\documentclass[aps,prl,twocolumn,showpacs]{revtex4}
\usepackage{amssymb}
\usepackage{graphicx}
\usepackage{amsmath}
\usepackage{amsfonts}

\begin{document}

\title{Luttinger theorem and low energy properties of ideal Haldane-Sutherland Liquids}
\date{\today}
\author{Tai-Kai Ng$^{1,2}$}
\affiliation{$^{1}$ Department of Physics, Hong Kong University of Science and Technology, Clear Water Bay, Hong Kong, China}
\affiliation{$^{2}$ Hong Kong Academy for Gifted Education, Shatin, Hong Kong, China}

\begin{abstract}
      We study in this paper the properties of a many body system of fermions obeying exclusion-statistics (Haldane liquid) where the origin of exclusion statistics is coming from an interaction-induced displacement field $\mathbf{a}_{\mathbf{k}}$ introduced by Sutherland (Haldane-Sutherland liquid). In particular we show how the Luttinger Theorem becomes compatible with exclusion statistics as a result of momentum conservation and adiabaticity. As a result, the low energy properties of Haldane-Sutherland liquids are Fermi/Luttinger liquid-like.

\end{abstract}

\maketitle

\textit{Introduction}---
   The notion of statistical interaction was first proposed by Haldane\cite{ex1}. Let the number of available single particle states for a many-body system of (spinless) fermions from momentum range $\mathbf{k}$ to $\mathbf{k}+d\mathbf{k}$ be $({dk\over2\pi})^d z(\mathbf{k})$  where $z(\mathbf{k})=1$ for normal fermions. Haldane suggested that the effects of interaction in many interacting systems can be captured by an interaction-induced renormalization of the fermion density of states which depends on the number distribution of all other particles\cite{ex1,ex2}($z(\mathbf{k})\rightarrow z(\mathbf{k};[n])\neq1$). This is the case for one-dimensional Bethe-Ansatz solvable models where exclusion statistics can be understood as the result of an interaction induced {\em displacement field} that shifts the momentum $k$ carried by particles\cite{ex1,exs,ex1d,su1}
        \begin{equation}
        \label{km1}
        k[n]=k_0+a_{k}[n],
        \end{equation}
         where $k$ and $k_0$ are the kinetic and canonical momentum of the particle, respectively and $a_{k}[n]$ is an interaction-induced displacement field acting on the particle with momentum $k$.

     For spinless fermions in one dimension, the displacement field $a_k$ has a form
     \begin{equation}
     \label{bethe}
     a_{k}[n]={1\over L}\sum_{k'}g(k-k')n_{k'}
     \end{equation}
     where $g(k-k')=-g(k'-k)$ represents scattering phase shift between particles\cite{su1}. The low energy physics of the system is characterized by the displacement field together with a free-particle Hamiltonian,
       \begin{equation}
       \label{hke}
       H[n]=\sum_{k}(\varepsilon_{k}-\mu)n_{k}.
       \end{equation}

       This particular form of displacement field $a$ and Hamiltonian has been generalized to dimensions $d>1$ phenomenologically by Sutherland, where $k(k_0)\rightarrow\mathbf{k}(\mathbf{k}_0)$ and $a_{k}[n]\rightarrow \mathbf{a}_{\mathbf{k}}[n]={1\over V}\sum_{\mathbf{k}'} \mathbf{g}(\mathbf{k}-\mathbf{k}')n_{\mathbf{k}'}$\cite{su1,su2} become $d$-dimensional vectors.
     The system is characterized by the same Hamiltonian\ (\ref{hke}). We shall call the resulting fermion liquid the {\em ideal} Haldane-Sutherland (HS) liquid in the following.

     In this paper we study the low energy properties of ideal HS liquids in arbitrary dimensions. In particular, we wish to investigate whether HS liquid can be a suitable model for strongly correlated electrons in dimensions $d>1$ and whether there exists any connection between HS liquids and Landau Fermi liquids\cite{lan}.  More specifically, we consider a system of spinless fermions where the kinetic momenta carried by fermions are shifted by an interaction-induced displacement field,
      \begin{subequations}
      \label{dfield}
      \begin{equation}
        \label{km2}
        \mathbf{k}[n]=\mathbf{k}_0+\mathbf{a}_{\mathbf{k}}[n],
        \end{equation}
        with
        \begin{equation}
      \label{aspin}
      \mathbf{a}_{k}[n]={1\over V}\sum_{\mathbf{k}'}\mathbf{g}_{\mathbf{k}\mathbf{k}'}n_{\mathbf{k}'},
      \end{equation}
      \end{subequations}
       where $\mathbf{g}_{\mathbf{k}\mathbf{k}'}$ is general, phenomenological function which may or may not be related to scattering phase shifts as in $1d$ Bethe-Ansatz solvable models. The Hamiltonian has kinetic energy term only as in\ (\ref{hke}) except $k\rightarrow\mathbf{k}$.

       The relation between the displacement field $\mathbf{a}_{\mathbf{k}}$ and exclusion statistics can be seen most easily by noting that the shift in kinetic momentum $\mathbf{k}_0\rightarrow\mathbf{k}$ results in a coordinate transformation of the system when the quasi-particles are labeled by kinetic momentum $\mathbf{k}$, where
       \begin{subequations}
       \begin{equation}
       \label{trans}
       {1\over V}\sum_{\mathbf{k}_0}=\int {d^dk_0\over(2\pi)^d}\rightarrow\int z(\mathbf{k};[n]){d^dk\over(2\pi)^d}= {1\over V}\sum_{\mathbf{k}}
       \end{equation}
       where $z(\mathbf{k};[n])=({\Delta^d\mathbf{k}_0\over\Delta^d\mathbf{k}})=Det|\mathbf{M}(\mathbf{k};[n])|$ is the Jacobian for the transformation between the coordinates $\mathbf{k}_0$ and $\mathbf{k}$,
       \begin{equation}
       \label{m}
        \mathbf{M}_{ij}(\mathbf{k};[n])={\partial k_{0i}\over\partial k_j}=\delta_{ij}-\partial_{k_{j}}a_{\mathbf{k}i}[n], (i,j=1,..,d)
        \end{equation}
        \end{subequations}
        is the transformation matrix\cite{su1}.
       The Jacobian $z(\mathbf{k};[n])$ describes the renormalization in the density of states (DOS) of the system when the phase space is transformed from canonical- to kinetic- momentum representation\cite{su1,Xiao}. The dependence of $z(\mathbf{k}[n])$ on occupation numbers $[n_{\mathbf{k}}]$ is precisely exclusion statistics.

        For ideal HS liquids, the ground state describes a filled Fermi sea with volume given by
      \begin{eqnarray}
       \label{q0}
       n={N\over V} & = & \int {d^dk\over(2\pi)^d}z(\mathbf{k};[n^{(0)}])n^{(0)}_{\mathbf{k}}  \\ \nonumber
       & = & \int {d^dk\over(2\pi)^d}z(\mathbf{k};[n^{(0)}])\theta(\varepsilon_F-\varepsilon_{\mathbf{k}})
       \end{eqnarray}
     where $\varepsilon_F$ is the Fermi energy. We note that the Fermi sea
     does not enclose a volume that satisfy Luttinger theorem in general when $z(\mathbf{k};[n])\neq1$\cite{ex1,ex2,iguchi}.

     As an example we consider the ideal HS liquid in one dimension with $\varepsilon_{k}=k^2/2m$ and $g_{kk'}=\pi\gamma sgn(k-k')$\cite{ex1d,su1,wu1}. In this case, it is easy to obtain from Eq.\ (\ref{bethe}),
     \begin{subequations}
     \label{fs1d}
     \begin{equation}
      z(k)={1\over1+\gamma n_k},
      \end{equation}
     and the Fermi volume is given by
     \begin{equation}
     V_F=2k_F=(1+\gamma)V_F^{(0)}=2(1+\gamma)k_F^{(0)},
     \end{equation}
     \end{subequations}
     where $k_F^{(0)}=\pi n$ is the Fermi wave-vector for spinless {\em free} fermions in $1d$. We note that this result violates the Luttinger theorem\cite{lut} which was believed to be satisfied for general interacting fermions in one dimension with ground states that respect translational symmetry\cite{os1,os2}. This is a key issue we shall address in the following. For convenience we shall also introduce the ``excluson" particle number $\tilde{n}_{\mathbf{k}}=z(\mathbf{k};[n])n_{\mathbf{k}}$\cite{ex2,wu1} in the following. $\tilde{n}_{\mathbf{k}}$ satisfies the inequality $0 \leq\tilde{n}_{\mathbf{k}}\leq z(\mathbf{k};[n])\neq1$ which is another way of expressing exclusion statistics\cite{ex2,ex1d,su1}. \\

\textit{Quasi-particles in Haldane-Sutherland liquid}---
    To understand the role of Luttinger theorem in ideal HS liquid we have to understand the properties of quasi-particles in HS liquid. We start by recalling quasi-particle properties in ordinary Fermi gas where a quasi-particle with momentum $\mathbf{k}$ is generated by adding/removing a particle to/from ground state. 

    In this case, the added particle can be represents as a change in particle distribution $\delta n_{\mathbf{k}'}={(2\pi)^d\over V}\delta^d(\mathbf{k}-\mathbf{k}')$. The added particle carries energy $\varepsilon_{\mathbf{k}}$ and momentum $\mathbf{k}$. The velocity of the quasi-particle is $\mathbf{v}_{\mathbf{k}}=\nabla_{\mathbf{k}}\varepsilon_{\mathbf{k}}$.

    The situation is rather different for HS liquids where the momentum carried by all other particles are shifted when a particle is added into the system.

     Let $\mathbf{k}=\mathbf{k}_G+\delta\mathbf{k}$ where $\mathbf{k}_G=\mathbf{k}_0+\mathbf{a}_{\mathbf{k}_G}[n^{(0)}]$ is the momentum for particle in state $\mathbf{k}$ when the system is in its ground state and $\delta\mathbf{k}$ is the momentum shift when an additional particle  $\delta \tilde{n}_{\mathbf{k}'}^{ex}={(2\pi)^d\over V}\delta^d(\mathbf{p}-\mathbf{k}')$ is introduced into the system. Then
       \begin{subequations}
       \begin{equation}
       \label{shift1}
       \Delta^d\mathbf{k}\tilde{n}_{\mathbf{k}}=\Delta^d\mathbf{k}_G\tilde{n}^{(0)}_{\mathbf{k}_G},
       \end{equation}
       (particle number conservation)\cite{Xiao} or
       \begin{equation}
       \label{shift2}
       \Delta^d\mathbf{k}\tilde{n}_{\mathbf{k}}=\Delta^d\mathbf{k}\tilde{z}(\mathbf{k})\tilde{n}^{(0)}_{\mathbf{k}-\delta\mathbf{k}},
       \end{equation}
       for $\mathbf{k}\neq\mathbf{p}$, where $\tilde{n}_{\mathbf{k}}$ and $\tilde{n}^{(0)}_{\mathbf{k}}$ are the excluson occupation numbers of the state $\mathbf{k}$  {\em after} and {\em before} the particle is added into the system, respectively, $\Delta^d\mathbf{k}$ and $\Delta^d\mathbf{k}_G$ are the corresponding phase space volume elements around momentum points $\mathbf{k}$ and $\mathbf{k}_G=\mathbf{k}-\delta\mathbf{k}$, respectively. $\tilde{z}(\mathbf{k})=({\Delta^d\mathbf{k}_G\over\Delta^d\mathbf{k}})=Det({\partial k_{Gi}\over\partial k_j})\sim1-\nabla_{\mathbf{k}}.\delta\mathbf{k}$ to first order in $\delta\mathbf{k}$, following Eq.\ (\ref{m}).

       Expanding Eq.\ (\ref{shift2}) to first order in $\delta\mathbf{k}$ we obtain
       \begin{equation}
       \label{shift3}
       \tilde{n}_{\mathbf{k}}=\tilde{n}^{(0)}_{\mathbf{k}}-(\nabla_{\mathbf{k}}.\delta\mathbf{k})\tilde{n}^{(0)}_{\mathbf{k}}
       -\delta\mathbf{k}.\nabla_{\mathbf{k}}\tilde{n}^{(0)}_{\mathbf{k}},
       \end{equation}
       \end{subequations}
       i.e., the shift in momentum results in a change in occupation numbers $\delta\tilde{n}_{\mathbf{k}}=-(\nabla_{\mathbf{k}}.\delta\mathbf{k}) \tilde{n}^{(0)}_{\mathbf{k}}-\delta\mathbf{k}.\nabla_{\mathbf{k}}\tilde{n}^{(0)}_{\mathbf{k}}$ for $\mathbf{k}\neq\mathbf{p}$. The added particle together with the induced changes $\delta\tilde{n}_{\mathbf{k}}$ defines a quasi-particle in HS liquid\cite{su1}.

       The total ``charge" carried by the quasi-particle is
       \begin{equation}
       \label{dq}
       \delta N= V\left(\int {d^dk\over(2\pi)^d}\left(\delta\tilde{n}^{ex}_{\mathbf{k}}+\tilde{n}_{\mathbf{k}}\right)- \int{d^dk_G\over(2\pi)^d}\tilde{n}_{\mathbf{k}_G}\right)=1,
       \end{equation}
       since shift in momentum doesn't create or destroy particles.

       Using Eq.\ (\ref{shift1}) it is also straightforward to show that the {\em total} momentum carried by a quasi-particle $(\mathbf{p})$ is given by
     \begin{eqnarray}
       \label{dp1}
       \mathbf{P}(\mathbf{p}) & = & \mathbf{p}+V\left(\int {d^dk\over(2\pi)^d}\mathbf{k}\tilde{n}_{\mathbf{k}}-{d^dk_G\over(2\pi)^d}\mathbf{k}_G\tilde{n}^{(0)}_{\mathbf{k}_G}\right)  \\ \nonumber
       & = & \mathbf{p}+V\int {d^dk_G\over(2\pi)^d}\delta\mathbf{k}\tilde{n}^{(0)}_{\mathbf{k}_G}.
       \end{eqnarray}
        We see that the total momentum carried by the quasi-particle is the momentum of the {\em bare} particle $+$ the total shift in momentum of the particles in the Fermi sea as a result of the displacement field.

        Similarly, the quasi-particle energy is
        \begin{eqnarray}
       \label{denergy}
       \tilde{\varepsilon}_{\mathbf{p}} & = & \varepsilon_{\mathbf{p}}+V\int {d^dk_G\over(2\pi)^d}[\varepsilon_{\mathbf{k}}-\varepsilon_{\mathbf{k}_G}]\tilde{n}^{(0)}_{\mathbf{k}_G}  \\ \nonumber
       & \sim & \varepsilon_{\mathbf{p}}+V\int {d^dk_G\over(2\pi)^d}[\delta\mathbf{k}.\nabla_{\mathbf{k}_G}\varepsilon_{\mathbf{k}_G}]\tilde{n}^{(0)}_{\mathbf{k}_G}
       \end{eqnarray}
       to first order in $\delta\mathbf{k}$. The velocity of the quasi-particle is $\mathbf{v}_{\mathbf{p}}=\nabla_{\mathbf{P}}\tilde{\varepsilon}_{\mathbf{p}}$.

        Next we determine $\delta\mathbf{k}$. Using Eq.\ (\ref{dfield}), we obtain
       \begin{eqnarray}
        \label{deltak1}
        \mathbf{k}_G+\delta\mathbf{k} & = & \int{d^dk_G'\over(2\pi)^d}\mathbf{g}_{\mathbf{k}_G+\delta\mathbf{k},\mathbf{k}_G'+\delta\mathbf{k}'}  \\ \nonumber
         & & \times(\tilde{n}_{\mathbf{k}_G'}^{(0)} +\delta\tilde{n}^{ex}_{\mathbf{k}_G'+\delta\mathbf{k}'}).
        \end{eqnarray}
        Expanding both sides of the equation to linear order in $\delta\mathbf{k}$, we obtain
        \begin{eqnarray}
        \label{deltak2}
        \delta\mathbf{k} & = & \int{d^dk_G'\over(2\pi)^d}\tilde{n}_{\mathbf{k}_G'}^{(0)}\left(\delta\mathbf{k}.\nabla_{\mathbf{k}_G}+\delta\mathbf{k}'.\nabla_{\mathbf{k}_G'}\right)\mathbf{g}_{\mathbf{k}_G\mathbf{k}_G'}
          \\ \nonumber
        & & +{1\over V}\mathbf{g}_{\mathbf{k}_G\mathbf{p}},
        \end{eqnarray}
        which is a linear equation for $\delta\mathbf{k}$.

        As an example, we consider again 1d spinless fermions with $\varepsilon_{k}=k^2/2m$ and $g(k-k')=\pi\gamma sgn(k-k')$\cite{su1,wu1}. It is easy to show that
        \begin{subequations}
        \begin{eqnarray}
       \label{dpf}
       P(p) & = & p+\int {dk_G\over 2\pi}g(k_G-p)\tilde{n}^{(0)}_{k_G}  \\ \nonumber
       & = & p_0
       \end{eqnarray}
       where $p_0$ is the canonical momentum of the particle and
       \begin{eqnarray}
       \label{denergyf}
       \tilde{\varepsilon}_{p}-\mu & = & {p^2\over2m}-{k_F^2\over2m}+\int {dk_G\over 2\pi}{k_G\over m}g(k_G-p)\tilde{n}^{(0)}_{k_G}  \\ \nonumber
       & = & {p^2\over2m}-{k_F^2\over2m}, \  (|p|>k_F)  \\ \nonumber
       & = & {1\over1+\gamma}\left({p^2\over2m}-{k_F^2\over2m}\right), \   (|p|<k_F)
       \end{eqnarray}
       where $k_F=(1+\gamma)k_F^{(0)}$.
       \end{subequations}

         Using Eq.\ (\ref{dfield}), it is also easy to show
         \begin{subequations}
         \label{pr0}
         \begin{eqnarray}
       \label{pr1}
       p & = & p_0+{\gamma k_F\over1+\gamma}sgn(p), \ (|p|>k_F)    \\ \nonumber
       & = & (1+\gamma)p_0. \   (|p|<k_F)
       \end{eqnarray}
          Notice that the quasi-particle velocity
       \begin{equation}
       \label{velocity}
       v_{p}={d\tilde{\varepsilon}_{p}\over dP}={d\tilde{\varepsilon}_{p}\over dp_0}={p\over m}
       \end{equation}
       \end{subequations}
         has no discontinuity across $|p|=k_F$\cite{su1,wu1}.

         We shall see in the following that the result $P(p)=p_0$ is not a coincidence but is a general property of Haldane-Sutherland liquids that respect translational symmetry.\\

 \textit{Momentum Conservation and Luttinger Theorem}---
          We consider a system of interacting fermions moving on a lattice with size $L^d$. Employing periodic boundary condition, the {\em total} momentum of the fermion liquid is quantized with
         \begin{equation}
         \label{Bloch}
         \mathbf{P}_{tot}={2\pi\over L} (n_1,n_2,...n_d)
         \end{equation}
         where $n_i$'s are integers. The quantization of {\em total} momentum is a result of lattice-translational symmetry of the system and is independent of fermion-fermion interaction.

           Next we consider HS liquids. In this case the total momentum of the system is given by
           \begin{subequations}
         \begin{eqnarray}
         \label{BLS}
         \mathbf{P}_{tot} & = & \int {d^dk\over(2\pi)^d}\mathbf{k}\tilde{n}_{\mathbf{k}}=\int {d^dk\over(2\pi)^d}\left(\mathbf{k}_0+\mathbf{a}_{\mathbf{k}}[n]\right)\tilde{n}_{\mathbf{k}}  \\ \nonumber
         & = & \mathbf{P}^{(0)}_{tot}+\mathbf{P}_c={2\pi\over L}(n_{1},n_{2},...n_{d}),
         \end{eqnarray}
          where
          \begin{eqnarray}
         \label{BLS1}
         \mathbf{P}^{(0)}_{tot} & = & \int {d^dk\over(2\pi)^d}\mathbf{k}_0\tilde{n}_{\mathbf{k}}=\int {d^dk_0\over(2\pi)^d}\mathbf{k}_0n_{\mathbf{k}}  \\ \nonumber
         & = & {2\pi\over L}\sum_i(l_{i1},l_{i2},...l_{id})
         \end{eqnarray}
         where $\mathbf{k}_{0i}={2\pi\over L}(l_{i1},l_{i2},..,l_{id})$ are the canonical momentum of individual occupied states $\mathbf{k}_i$, and
         \begin{equation}
         \label{BLSc}
         \mathbf{P}_{c}=\int {d^dk\over(2\pi)^d}\mathbf{a}_{\mathbf{k}}[n]\tilde{n}_{\mathbf{k}}={2\pi\over L}(m_{1},m_{2},...m_{d})
         \end{equation}
         is a correction term coming from the displacement field $\mathbf{a}_{\mathbf{k}}$, $m_i=n_i-\sum_jl_{ji}$.
         \end{subequations}

          In particular, as the displacement fields are results of fermion-fermion interaction, it is expected that $\mathbf{P}_c=0$ and $\mathbf{P}_{tot}=\mathbf{P}^{(0)}_{tot}$ if the total momentum of the system remains invariant when the interaction is turned on adiabatically. This is valid as long as the fermion-fermion interaction term respects translational symmetry and there is no spontaneous symmetry breaking during the process when the fermion-fermion interaction is turned on.

          For HS liquids, the condition $\mathbf{P}_c=0$ implies
          \[
          \int {d^dk\over(2\pi)^d}\int {d^dk'\over(2\pi)^d}\tilde{n}_{\mathbf{k}}\mathbf{g}_{\mathbf{k}\mathbf{k}'}\tilde{n}_{\mathbf{k}'}=0, \]
           and $\mathbf{g}_{\mathbf{k}\mathbf{k}'}=-\mathbf{g}_{\mathbf{k}'\mathbf{k}}$, which is satisfied by Bethe-Ansatz solvable spinless fermion models in one dimension (Eq.\ (\ref{bethe}))\cite{su1,wu1}.

          As a result, in the case of adding one particle into the system with zero initial total momentum $\mathbf{P}_{tot}=0$, the {\em total} momentum carried by the added particle is
          \begin{eqnarray}
       \label{pquasi}
       \mathbf{P}(\mathbf{p})  & = & \mathbf{p}+\int {d^dk\over(2\pi)^d}\delta\mathbf{k}\tilde{n}^{(0)}_{\mathbf{k}}  \\ \nonumber
        & = & \mathbf{p}+\int {d^dk\over(2\pi)^d}\mathbf{g}_{\mathbf{k}\mathbf{p}}\tilde{n}^{(0)}_{\mathbf{k}}  \\ \nonumber
       & = & \mathbf{p}_0.
       \end{eqnarray}
       We have used Eqs.\ (\ref{dfield}) and\ (\ref{deltak2}) in deriving the above result.

        We now consider an excitation in a one dimensional HS liquid where a particle with momentum $-k_F$ is moved to momentum $k_F$ or vice versa, i.e. we move the particle from one side of the Fermi surface to another. This is a zero-energy excitation with momentum transfer $\pm 2k_F^{(0)}$ but not $\pm 2k_F$ because the {\em total momentum} carried by a quasi-particle is $\pm k_F^{(0)}$. This result is in agreement with the proof of the Luttinger Theorem by Yamanaka {et al.} in one dimension\cite{os1}, where they prove that a fermion system always has zero energy excitation with momentum transfer $2k_F^{(0)}$ if the system does not break translational symmetry. The argument can be extended rather straightforwardly to dimensions $d>1$. In this case we can construct zero energy excitations by moving fermions from one part of the Fermi surface with Fermi momentum $\mathbf{k}_F$ to another with Fermi momentum $\mathbf{p}_F$, the total momentum transfer is $\mathbf{p}_F^{(0)}-\mathbf{k}_F^{(0)}$ (not $\mathbf{p}_F-\mathbf{k}_F$), in agreement with Oshikawa's proof of Luttinger theorem at $d>1$\cite{os2}.

        It should be noted that the ``proof" of Luttinger theorem here (and in refs.\cite{os1,os2}) does not measure the Fermi sea volume directly which is what the Luttinger Theorem is supposed to address\cite{lut}. 
        The distinction between the Luttinger theorem (measured by momentum transfer) and ``real" Fermi sea volume can be seen by an example of spinless fermions on a $1d$ lattice with $\varepsilon_{k}=-2t\cos(k)$ and $g(k-k')=\pi\gamma sgn(k-k')$\cite{su1,wu1}. In this case the Fermi volume is given by Eq.\ (\ref{fs1d}). The fermion band is filled and the system becomes {\em insulating} when $k_F=(1+\gamma)k_F^{(0)}=\pi$ (assuming $\gamma>0$). Nevertheless the momentum transfer from moving a fermion from one side of Fermi surface to the other side remains to be $2k_F^{(0)}$, independent of the exclusion parameter $\gamma$, i.e. we observe that the Fermi sea volume measured by Luttinger Theorem is {\em not} renormalized by fermion-fermion interaction all the way up to $(1+\gamma)k_F^{(0)}=\pi$ when the system becomes insulating, in agreement with Fermi liquid theory\cite{mit}. The approach of the system towards insulating state is indicated by the ``real" Fermi sea volume $2k_F$, but not by the Fermi sea volume according to the Luttinger theorem $=2k_F^{(0)}$. \\

\textit{Effective Fermi liquid description and spin-$1/2$ fermions}---
        The above ``proof" of the Luttinger theorem in ideal HS liquids suggests that it is possible to construct an effective Landau Fermi liquid-type theory (or Luttinger Liquid theory at one 1d) describing the low-energy quasi-particle dynamics in HS liquids. The Fermi/Luttinger liquid satisfies Luttinger Theorem in the sense we describe above.

        For ideal HS liquids, the construction of effective Landau/Luttinger liquid theory is straightforward in principle. There is no scattering between quasi-particles and the quasi-particles occupation number represent exact eigenstates of the system. As a result, the energy of the system is a functional of quasi-particle occupation numbers $n$, $E=E[n]$ and the (semi-classical) low energy dynamics can be determined by expanding $E[n]=E[n^{(0)}+\delta n]$ to second order in $\delta n$. $n^{(0)}$ being the ground state quasi-particle occupation numbers, following the argument of Landau\cite{lan}. The calculation is complicated by the fact that we have to take into account the shifts in momentum, and correspondingly the quasi-particle energy as $\delta n$ changes.

         This calculation has been done by Wu {\em et al.}\cite{wu1} for $1d$ spinless fermion systems where they confirmed that $1d$ HS liquids are Luttinger liquids. The construction of the effective Landau energy functional and corresponding transport equation can also be implemented for dimensions $d>1$. The calculation is straightforward but tedious. We shall report these details in a separate paper.


          Lastly we make a few comments about spin-$1/2$ HS liquids here. The Bethe Ansatz solution for spin-$1/2$ interacting fermions at $1d$ results in a low energy effective theory with spin-charge separation\cite{lw}. This state can be described as an effective 2-components HS liquid with the 2 components being spin (s) and charge (c)\cite{wu2,ov}. The charge and spin ``quasi-particles" are characterized by quasi-momentum $k$'s and rapidities $\lambda$'s, respectively which are determined by displacement fields $a^{c(s)}$ that depend on quasi-particle occupation numbers $n^s_{\lambda}$ and $n^c_{k}$,
          \[
              a^{\alpha}_p={1\over L}\sum_{\beta p'}g^{\alpha\beta}_{p,p'}n_{p'}^{\beta}  \]
              where $\alpha,\beta=c,s$ and $p(p')=k,\lambda$ for $\alpha(\beta)=c,s$\cite{wu2}. This description may not be applicable to fermion systems at dimensions $d>1$ where spin-charge separation does not take place.

              We propose here a simple, phenomenological description of spin-$1/2$ ideal HS liquids at $d>1$ with no spin-charge separation. The quasi-particles in this HS state carries both spin $\sigma=\pm{1\over2}$ and kinetic momentum $\mathbf{k}_{\sigma}$. The momentum $\mathbf{k}_{\sigma}$ is determined by a spin-dependent displacement field
       \begin{subequations}
      \label{dfields}
      \begin{equation}
        \label{kspin}
        \mathbf{k}_{\sigma}[n]=\mathbf{k}_{0\sigma}+\mathbf{a}_{\mathbf{k}\sigma}[n],
        \end{equation}
        where $\mathbf{k}_{0\sigma}$ is the canonical momentum and
        \begin{equation}
      \label{aspin}
      \mathbf{a}_{k\sigma}[n]={1\over V}\sum_{\mathbf{k}'\sigma'}\mathbf{g}_{\mathbf{k}\sigma;\mathbf{k}'\sigma'}n_{\mathbf{k}'\sigma'}
      \end{equation}
      is the generalized spin-dependent displacement field.
      \end{subequations}

        Following analysis for spinless fermions we can show that the resulting quasi-particles carry spin-$1/2$ and unit charge. The system satisfies Luttinger theorem (in the sense of momentum transfer) if $\mathbf{g}_{\mathbf{k}\sigma;\mathbf{k}'\sigma'}=-\mathbf{g}_{\mathbf{k}'\sigma';\mathbf{k}\sigma}$ and connects to Fermi liquid theory as we described above.\\

      \textit{Summary}---
    In this paper we study the properties of ideal Haldane-Sutherland liquids at arbitrary dimensions and demonstrate that momentum conservation and adiabaticity leads to a proof of Luttinger theorem for Haldane-Sutherland liquids. The Luttinger theorem proved this way relies on measuring the momenta carried by quasi-particles and does not measure directly the volume of underlying Fermi sea. It should be noted that the key identity $\mathbf{g}_{\mathbf{k}\mathbf{k}'}=-\mathbf{g}_{\mathbf{k}'\mathbf{k}}$ we derive in this paper (and its generalization to spin-$1/2$ systems\cite{lw}) has been known for $1d$ Bethe-Ansatz solvable models. The relation of this identity to momentum conservation and adiabaticity in HS liquids is the new finding in this paper. The theorem connects HS liquid to usual Fermi liquids and implies that the low energy properties of HS liquids are Fermi liquid-like (or Luttinger liquid-like in $1d$). It provides a plausible microscopic mechanism of how the Mott metal-insulator transition may occur in fermion systems and suggests that HS liquid may be a good starting point modeling strongly correlated electron systems.\\

  \subsubsection{acknowledgement}

           This work is supported by HKRGC through grant No. HKUST3/CRF/13G.



\end{document}